\documentclass[epj]{svjour}
\usepackage{graphics}
\begin{document}
\title{Progress on reactions with exotic nuclei}
\author{F.M. Nunes\inst{1}\thanks{Conference presenter}, 
A.M. Moro\inst{2}, A.M. Mukhamedzhanov\inst{3}  and N.C. Summers\inst{1}
\thanks{This work has been partially
supported by National Superconducting Cyclotron Laboratory at 
Michigan State University, the Portuguese Foundation for Science (F.C.T.), 
under the grant POCTIC/36282/99,  the Department of Energy
under Grant No.\@ DE-FG03-93ER40773 and the U.\,S. National Science
Foundation under Grant No.\ PHY-0140343.}
}                     % Do not remove
\authorrunning{F.M. Nunes et al.}
\institute{
NSCL and Department of Physics and Astronomy,  MSU, East Lansing MI 48824,  USA.
\and 
Departamento de FAMN, Universidad de Sevilla, Aptdo. 1065, 41080 Sevilla, Spain.
\and
Cyclotron Institute, Texas A\& M University, College Station TX 77843 USA.
}
\date{Received: 17th November 2004 / Revised version: 15th April 2005}
% The correct dates will be entered by Springer
%
\abstract{
Modelling breakup reactions with exotic nuclei represents a challenge 
in several
ways. The CDCC method (continuum discretized coupled channel)
has been very successful in its various applications.  
Here, we briefly mention a few developments that have contributed 
to the progress in this field as well as some pertinent problems 
that remain to be answered.
\PACS{24.10.Ht, 24.10.Eq, 25.55.Hp, 25.70De, 27.20.+n}
} %end of abstract
\maketitle
\section{Introduction}
\label{intro}
\vspace{-0.2cm}

Light nuclei on the drip lines can be studied
through a variety of reactions. Models
for nuclear reactions have been developed in recent
years in order to incorporate the exotic features
of these dripline nuclei. The real challenge for
reaction theory lies in the low energy regime
where most approximations are not valid \cite{jpg}.

Three body effects  need to be
carefully considered in the lower energy regime.
At energies close to the breakup threshold, Integral Faddeev
Equations would be the appropriate choice. However, 
due to technical difficulties, the Continuum Discretized 
Coupled Channel Method (CDCC) \cite{cdcc} is the best working 
alternative. Here we consider specific features of breakup within
CDCC, namely the continuum couplings in the usual
breakup basis (section 2), and the alternative breakup mechanism
consisting of transfer to the continuum of the target (section 3).
Finally, in section 4, we briefly comment on remaining problems.

\section{Continuum couplings}
\label{ccbins}
\vspace{-0.2cm}

Measurements of $^8$B breakup are of 
importance to nuclear astrophysics. There have been several
experiments performed at different facilities to provide the
needed information on the $S_{17}$. Using our best understanding 
of the reaction mechanism, we assume the projectile can
be represented by  $^7$Be(inert)$+p$.
Within CDCC, the scattering states are binned up in energy
(or momentum) labelled by an index $\alpha$.
When the projectile breaks up through the interaction
with the target it can rearrange itself within the continuum.
The relevant  couplings, connect two continuum bins and have the form 
\begin{eqnarray} \label{eqcgs}
V_{\alpha;\alpha'}({\rm {\bf R}}) & = & < \phi_\alpha({\rm {\bf r}})|
V_{cT} ({\bf R _ c}) + V_{fT}({\bf R _ f})|\phi_{\alpha'}({\rm {\bf r}})>\;
\end{eqnarray}
where $r$ is the projectile internal relative motion ($c+f$),
$R$ is the relative motion between the projectile and the target
and $R_c$($R_f$) is the vector connecting the center
of mass of the core(fragment) to the center of mass of the target.
\begin{figure}
\resizebox{0.33\textwidth}{!}{\includegraphics{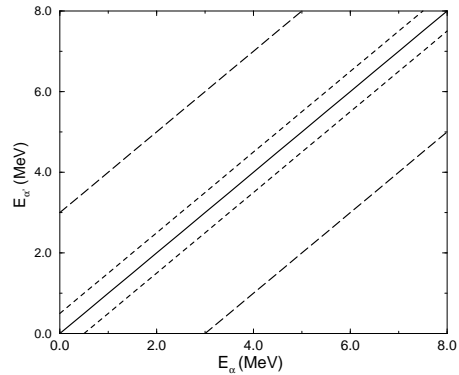}}
\caption{Representation of the relevant bins that need
to be considered in a CDCC calculation: for a monopole transitions 
(solid), dipole transitions (dashed) and hexadecapole
transitions (long-dashed). More details can be found in the text. }
\label{fig:1}       % Give a unique label
\end{figure}

The proximity to the breakup threshold has been shown to have 
important effects
in the reaction mechanism. For instance, in the $^8$B breakup 
around the Coulomb barrier  \cite{nunes99} Coulomb multistep effects 
reduced the cross
section up to 20\% but the most remarkable effect was
related to the nuclear peak at larger angles which disappeared
through continuum-continuum couplings.
Continuum couplings are a way of
looking into the effect of the final state
interaction, integral part of CDCC. 

The properties of these continuum couplings
and the influence they can have on breakup observables 
have been the object of a recent study \cite{nunes04}.
Their long range behaviour is preserved throughout the
multipole expansion, which slows down convergence:
these couplings are shown to behave as $1/R^2$ for dipole transitions
and $1/R^3$ for all higher multipoles. 

It was also found that continuum-continuum couplings have
certain patterns with core-fragment relative energy and relative
angular momentum. 
Monopole couplings are strongest
when the initial and final relative energies are the same
(represented in Fig. \ref{fig:1} by the solid line), a simple
consequence of the normalization of the bin wavefunction.
Dipole couplings are strongest when these energies differ
by an amount comparable to the energy width of the bin,
(represented in Fig. \ref{fig:1} by the area in between the dashed lines). 
The higher the order of the couplings, the larger the region
that needs to be taken into account. Tests on using
this property for optimizing the large CDCC calculations
have been performed. Optimization of lower partial waves
is of little interest since the number of bins involved are
typically small. It is for the larger partial waves ($l > 2$)
that calculations become heavy. Our tests show that couplings with
initial and final energies differing by several energy steps 
need to be considered  in order to get convergence.
This does not allow for significant improvement of the size and the
time of the calculations.

\section{Transfer to the continuum}
\vspace{-0.2cm}

\begin{figure}
\resizebox{0.4\textwidth}{!}{\includegraphics{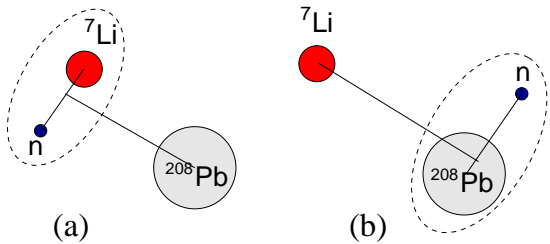}}
\caption{Schematic diagram for the breakup of $^8$Li on
$^{208}$Pb: (a) the standard breakup and (b) the
transfer to the continuum of the target. }
\label{fig:2}       % Give a unique label
\vspace{-0.5cm}
\end{figure}

A variety of breakup models are presently in use and, when two different 
models are applied to the same problem, there is
often a disparity in the predictions. 
In this sense, a generalized effort to bridge the various approaches 
is very much needed. One of the important issues lies in the choice
of the coordinate representation of the continuum wavefunctions.

We present results of a comparative study between the standard
CDCC breakup approach and the so called transfer 
to the continuum \cite{moro04}. In the standard breakup approach,
a projectile fragment is excited into the continuum, whilst keeping
the correlation to the projectile core. There
are cases where the correlation of the fragment with the target is more
important, and then an expansion in terms of the standard
breakup basis does not enable convergence within practical
limits. Such was the case for the breakup studies of $^6$He \cite{he6nd}
and $^8$Li \cite{li8moro} at energies around the Coulomb barrier.
In Fig. \ref{fig:2} we show the coordinate representation
for the breakup of $^8$Li on $^{208}$Pb in the standard approach (a)
and in the transfer to the continuum of the target approach (b).
The differences between the two approaches are 
over-emphasized when resonances (in any particular channel) 
play a role in the dissociation process.

Typically, with an option of using an expansion based on
the continuum of the projectile or the continuum of the target, 
one chooses the continuum of the more loosely bound
nucleus, since it will be more prone
to breaking up. However, there are some cases where this
choice is not clear. For example, in the $^7$Be(d,n)$^8$B reaction
\cite{be7dn} one can immediately expect the $^8$B continuum to be
very important given the binding energy $0.137$ MeV.
However, the deuteron breakup is often very strong too.
The inclusion of both continua, in the entrance channel and the
exit channel raise some orthogonality issues that need
to be addressed soon.

\section{Remaining problems}
\label{problems}
\vspace{-0.2cm}

Even though the $^8$B breakup application of CDCC has been 
extremely successful \cite{tostevin01}, low energy Notre Dame data and 
high energy NSCL/MSU data show  a $60$\% inconsistency in the 
quadrupole excitation strength. This is an extremely severe problem 
from the point of view  of direct capture \cite{b8neil}. Independently, 
accurate measurements have shown that
$^7$Be first excited state contributes to the ground state
of $^8$B \cite{b8gsi}. It is possible that core excitation
will help solve the puzzle.

Major advances have been performed on microscopic approaches 
to reactions which include the treatment of one and two particle
continuum (e.g. the Shell model embedded in the continuum model
\cite{smec}). 
Although the variety of reactions that can be addressed
through these microscopic models is rather limited, core
excitation is much better treated than within the 
CDCC few body approach.

\end{document}